\documentclass[a4paper]{spie}  
\usepackage[dvips]{graphicx,color}
\usepackage{subfigure}
\newcommand{\unit}[1]{\,\,\mathrm{#1}}

\title{Quantum dots as handles for optical manipulation } 

\author{Liselotte Jauffred\supit{1}, Marit Sletmoen\supit{2}, Fabian Czerwinski\supit{1},  and Lene Oddershede\supit{1}
\skiplinehalf
\supit{1}The Niels Bohr Institute, University of Copenhagen, Denmark
\skiplinehalf
\supit{2}NTNU, Trondheim, Norway}

\authorinfo{Further author information:\\
L.J.: E-mail: jauffred@nbi.dk\\
M.S.: E-mail: marit.sletmoen@ntnu.no\\
F.C.: E-mail: czerwinski@nbi.dk\\
L.O.: E-mail: odder@nbi.dk}

\begin{document} 
  \maketitle 

\begin{abstract}
Individual colloidal quantum dots can be optically trapped and manipulated by a single infrared
laser beam operated at low laser powers \cite{jauffred2008,jauffred2010}. If the absorption spectrum
and the emission wavelength of the trapping laser are appropriately chosen, the trapping laser light 
can act as a source for two-photon excitation of the trapped quantum dot. This eliminates the
need for an additional excitation laser in experiments where individual quantum dots
are used both as force transducers
and for visualization of the system. To use quantum dots as handles for quantitative optical 
force transduction, it is crucial to perform a precise force calibration. Here, we present an
Allan variance analysis \cite{fabian2009} 
of individual optically trapped quantum dots and show that the
optimal measurement time for experiments involving individual quantum dots is on the order of 0.3 seconds. 
Due to their small size and strong illumination,
quantum dots are optimal for single molecule assays where, optimally, the
presence of the tracer particle should not dominate the dynamics of the system. As an example, 
we investigated the thermal fluctuations of a DNA tether
using an individual colloidal quantum dot as marker, this being the smallest tracer for
tethered particle method reported.
\end{abstract}


\section{Introduction}
Colloidal quantum dots (QDs) are semiconducting
crystals with sizes in the range from a few to hundred of nanometers \cite{michalet2001,dubertret2002}. They are bright and
photo-stable with a broad excitation spectrum and a narrow emission
spectrum, normally distributed around a specific wavelength $\lambda$. Because of the high quantum yield and low bleaching, QDs have a broad range of applications in the
investigation of biological systems\cite{chan1998,michalet2001,michalet2005,larson2003}, e.g, single-particle tracking
of individual receptors in a cell membrane \cite{dahan2003} and in
vivo imaging \cite{ballou2004}.

Optical tweezers are often used to trap and manipulate nanometer-sized particles, the interest
arises from the fact that the nanoparticles serve as excellent handles for investigations
of individual biomolecules. Within the recent years there have been several reports on optical trapping
of metallic nano-particles \cite{Svoboda1994,bosanac2008,hansen2005,selhuber-unkel2008,haji2010} with
dimensions down to 8 nm. QD are among the nanoparticles with
inducible dipole moments which can be optically trapped by an individual infrared relatively 
weak laserbeam \cite{jauffred2008,jauffred2010}. From knowledge of the thermal fluctuations of
the nanoparticle within the trap, one can deduce the trapping force acting on the quantum dot
and the polarization of the particle \cite{jauffred2008}.

QDs can be exited by two-photon absorption of 
infrared laser light which simultaneously
traps the QDs. The absorption does not alter the trapping properties, e.g., the spring
constant, in any pronounced way 
\cite{jauffred2010}. Two-photon absorption of the trapping laser light does give rise
to  bleaching of QDs. However, bleaching is more rapid if the QDs in addition to
the trapping laser are illuminated by a Hg lamp \cite{jauffred2010}. Hence, it is beneficial
to only use a single light source for both trapping and excitation.

It is custom to perform force calibrations in order to quantitatively 
measure forces present in trapping assays. One common way to do this is through power spectral
analysis \cite{Berg-Sorensen2004}. This can be combined with Allan variance
analysis which provides information concerning the optimal time window to perform force
calibrations and system measurements, and which also serves as a method to minimize force
contributions in the experimental setting \cite{fabian2009}.

Here, we show how the optical 
trapping properties of individual QDs vary as a function of QD emission wavelength,
the emission wavelength being closely related to the physical size of the QD. Also, we
perform an Allan variance analysis of the time series originating from the optical trapping
experiment, this analysis revealing the optimal calibration window and quality of the experimental
setting. Finally, we present an example of how to use individual QDs as markers of single molecule
systems. We used QDs were as tracer particles of DNA tethered particle method (TPM) 
and the analysis revealed that
this small tracer particle had a minimum of influence on the dynamics of the DNA tether.


\section{Dependence of trapping strength on quantum dot size}\label{sec:QDsize}

\begin{figure}
\centering
\subfigure[\label{fig:TEM}TEM
pictures of a$\left.\right)$ 525 nm QDs, b$\left.\right)$ 585 nm QDs, c$\left.\right)$ 605 nm QDs, d$\left.\right)$ 655 nm QDs (note the elongated shape),
e$\left.\right)$ 705 nm QDs, f$\left.\right)$ 800 nm QDs.
The scale bars corresponds to 10 nm.]{\includegraphics[width=6.5cm]
{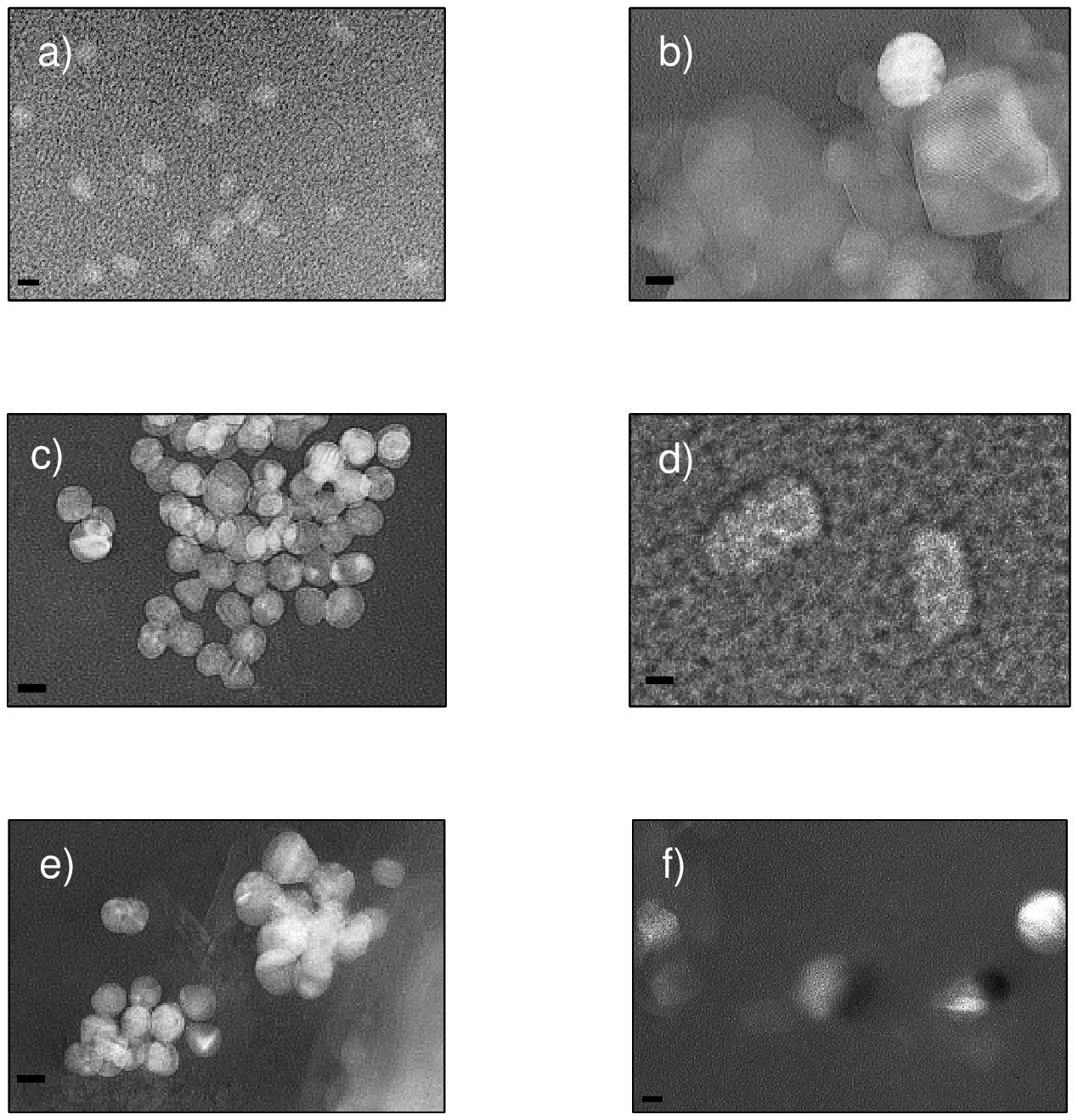}}\hspace{0.5cm}
\subfigure[\label{fig:powerSpectrum} Power spectrum of the lateral positions visited by an optically trapped QD (wavelength of 585 nm).
Full line is a Lorentzian fit to the data \cite{Hansen2006(pm)},
punctuated lines represents STD. The corner frequency $f_c$ of the Lorentzian fit is (159.3 $\pm$ 6.5) Hz. The inset shows the
distribution of positions visited by the QD in the optical trap for the same time
series, full line is a Gaussian fit.]{\includegraphics[width=10cm]
{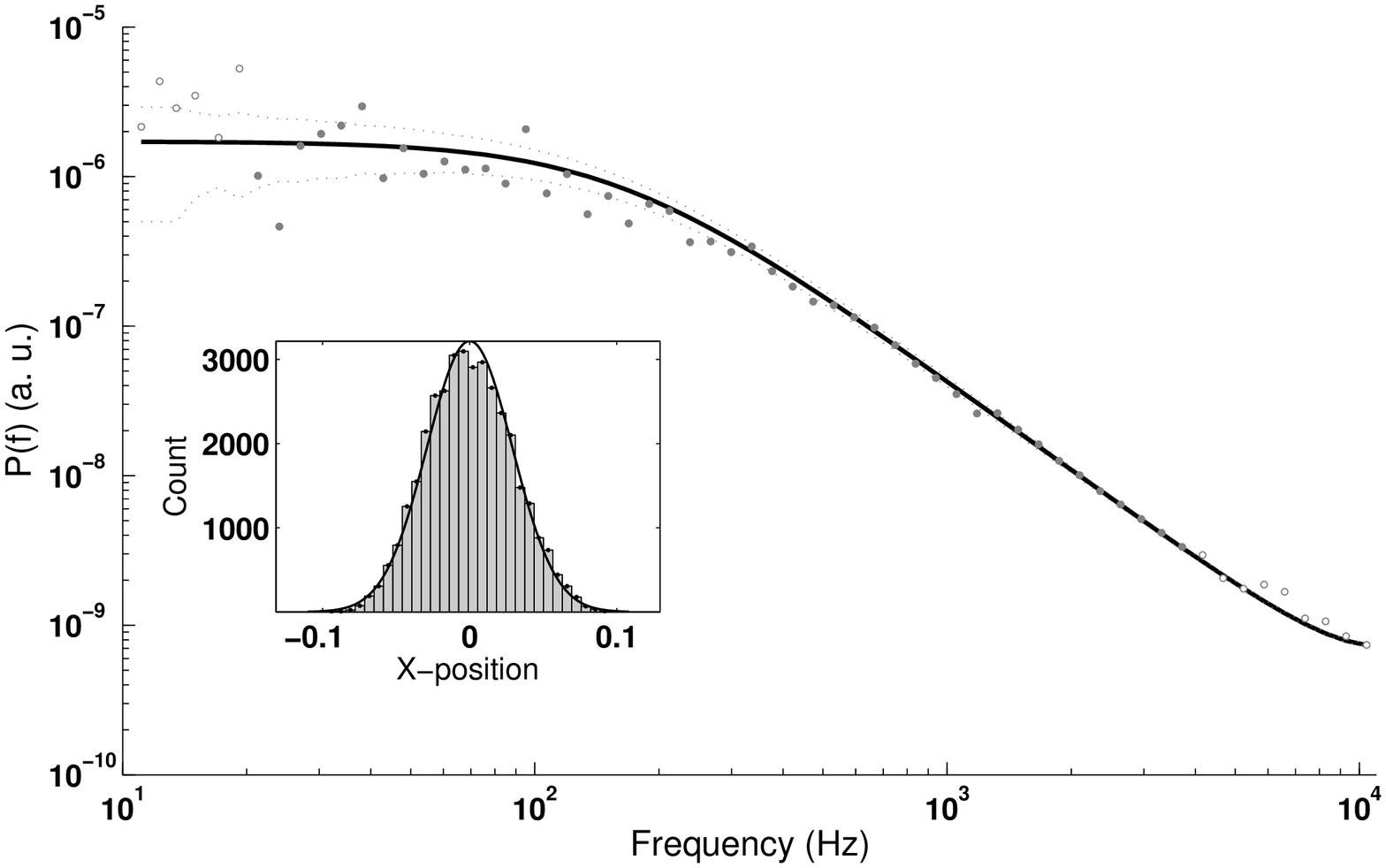}}
\label{fig:subfigureExample}\caption{}
\end{figure}

It is important to correctly choose the QD such that it matches the
given experimental goals and conditions, e.g., available excitation
lasers and filter cubes. Therefore, the optical trapping properties of QDs were determined as a 
function of emission wavelength
or size. The optical trap was based on a 1064 CW laser implemented in an inverted Leica SP5 microscope,
details regarding the experimental settings are given in References \cite{jauffred2008,jauffred2010}.
Figure \ref{fig:TEM} shows TEM images of the QDs investigated in the present study.

It is well established that the optical tweezers
exert a harmonic force on the trapped particle:
$\textbf{F}=-\kappa \textbf{x}$, where $\kappa$
denotes the trap stiffness and $\textbf{x}$ is the position
of the particle with respect to the center of the trap. To quantify
$\kappa$ and hence find the range of optical forces exertable on a
QD we performed a power spectral analysis of the time series using
the routines described in Reference \cite{Hansen2006(pm)}. The distance between
the QD and the cover slide surface was very large in comparison to the
radius of the QD, and we approximated the overall shape of the QD as
a sphere. Hence, the drag coefficient, $\gamma$, can be found by Stoke's
Law, $\gamma=6 \pi \eta d$, where $d$ is the effective diameter of the
QD.
The
equation of motion of a particle performing Brownian fluctuations
inside an optical trap at low Reynolds number in one direction is: 
\begin{equation}
\textbf{F}(t)=\gamma\frac{d\textbf{x}}{dt}+\kappa\textbf{x},
\end{equation}
where \textbf{F}(t) is a stochastic force resulting from the thermal motion of the media surrounding the 
trapped particle.
Fourier
transformation of the equation of motion gives the positional
power spectrum which follows a Lorentzian function: 
\begin{equation}
P(f)=\frac{k_BT}{\gamma \pi^2 (f^2+f_{c}^{2})},
\end{equation}
where $\gamma$ is the drag coefficient and the corner frequency is denoted $f_c$.
The corner frequency distinguishes 
the plateau region of slow fluctuations and the region with a scaling exponent of $-2$. This is the characteristic of Brownian motion for rapid fluctuations. $f_c$ is related to $\kappa$ and to $\gamma$ 
of the QD: $f_c = \frac{\kappa}{2 \pi \gamma}$.
Fig. \ref{fig:powerSpectrum} shows an example of the power spectrum of positions visited
by a 585 nm QD in a direction orthogonal to the propagating laser light. The full line is a
fit by a Lorentzian function taking into account aliasing and the filtering effects of the
quadrant photo-diode system \cite{Berg-Sorensen2003,Hansen2006(pm)}. 
The inset in Fig. \ref{fig:powerSpectrum} shows that positions visited by
the QD inside the optical trap follow a Gaussian distribution (full line).

\begin{table}
\centering
\begin{tabular}{|l|l|l|l|}
  \hline
$\lambda$ & core (nm) & composition& $d$ (nm)\\
  \hline
  525 nm & 3-4 & CdSe/ZnS &  $10\pm 2$ \\
  585 nm  & 5.3 & CdSe/ZnS & $26\pm 12$ \\
  605 nm  & (4$\times$9.4) & CdSe/ZnS & $13\pm 1$ \\
  655 nm & (6$\times$12) & CdSe/ZnS & ($40\pm 5$)$\times$($24\pm 3$) \\
  705 nm & ND & CdSeTe/ZnS & $16\pm 3$ \\
  800 nm & ND & CdSeTe/ZnS & $21\pm 3$ \\
  \hline
\end{tabular}\vspace{0.3cm}
\caption{Physical characteristics of investigated QDs. First column gives emission wavelength $\lambda$,
second column the size of the core which is given by the diameter if the QDs are spherical or by the
semi-major and semi-minor axes where the QDs are more ellipsoidal, third column
states the material composition \cite{kraus2008}. 
The last column gives the outer diameter $d$ as measured by TEM.}
\label{tabel}
\end{table}

Table \ref{tabel} gives an overview of the physical characteristics of the QDs used in the present
study. The QDs consist of a CdSe or CdSeTe core surrounded by a ZnS shell. 
All QDs were bought from Invitrogen who also
provided the informations shown in the 3 first rows in Table \ref{tabel}.
We made transmission electron microscopy (TEM) pictures of the QDs, examples are shown in Fig.
\ref{fig:TEM}, to determine the sizes.
  
\begin{figure}
  \centering
  \includegraphics[width=14cm]{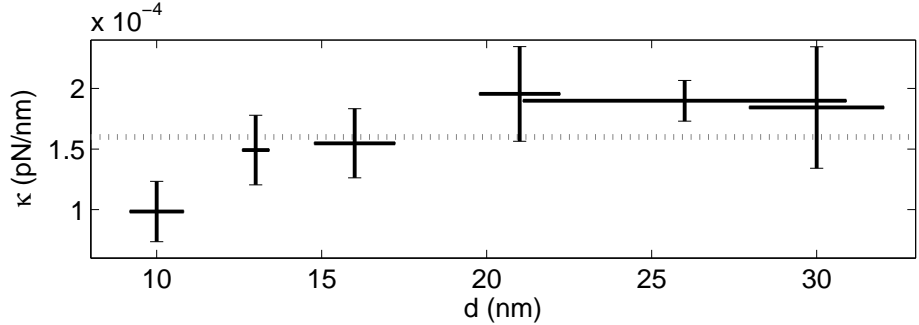}
  \caption {\label{fig:kappa4} Trapping strength $\kappa$ dependence on
diameter $d$ for laser power of 0.4 W. The points denotes the mean of the data
points calculated using the $d$ and $f_c$. In the case of the elongated 655 nm QD, $d\sim$ 30 nm. The dotted line corresponds to the mean value of $\kappa$. All error bars denote one SEM.}
  \end{figure} 

Fig. \ref{fig:kappa4}
shows the trap stiffness, $\kappa$, as a function of QD diameter $d$ as determined by the TEM pictures.
The value of $\kappa$ is calculated from the experimentally found $f_c$'s and the diameter, $d$. The dotted line
shows the average value: $\kappa=(1.6\pm0.4)\times10^{-4}$ pN/nm (mean $\pm$ STD).
For the investigated QDs, the trap stiffness appears constant, independent of emission wavelength, $\lambda$,
or physical size, $d$. However, a close inspection of Fig. \ref{fig:kappa4} indicates that $\kappa$ could be 
increasing as a function of QD size. This is supported by a Students t-test, which shows that $\kappa$ for 
the 525 QD is significantly smaller than $\kappa$ for the 800 nm QD (p = 0.6981).

\section{Allan variance analysis for quantum dots}
Allan variance analysis is a method to quantify noise. In comparison to power spectral analysis it is
in particular well suited to pinpoint low-frequency noise. In contrast to the normal variance, Allan
variance converges for most of the naturally occuring types of noise. Through Allan variance analysis one
can determine the optimal time for a measurement or calibraion, this is the time where the advantage of
drawing additional points from a Gaussian position distribution is overcome by the drift inherently
present in a real experiment. For a timeseries of $N$ elements and a total measurement time of 
$t_\textnormal{acq}=f_\textnormal{acq}^{-1}N$ the Allan variance, $\sigma_x^2\left(\tau\right)$, 
is defined as follows \cite{allan1966,fabian2009}:
\begin{equation}
\sigma_x^2\left(\tau\right)=\frac{1}{2}\left< \left(x_{i+1}-x_i\right)^2 \right>_\tau.
\end{equation}
Here, $x_i$ is the mean of a time interval of length $\tau=f_\textnormal{acq}^{-1}m$, $m$ being
the number of elements in this particular interval. In words, the Allan variance is half 
the averaged squared mean of neighboring intervals each of length $\tau$.

\begin{figure}
  \centering
  \includegraphics[width=12cm]{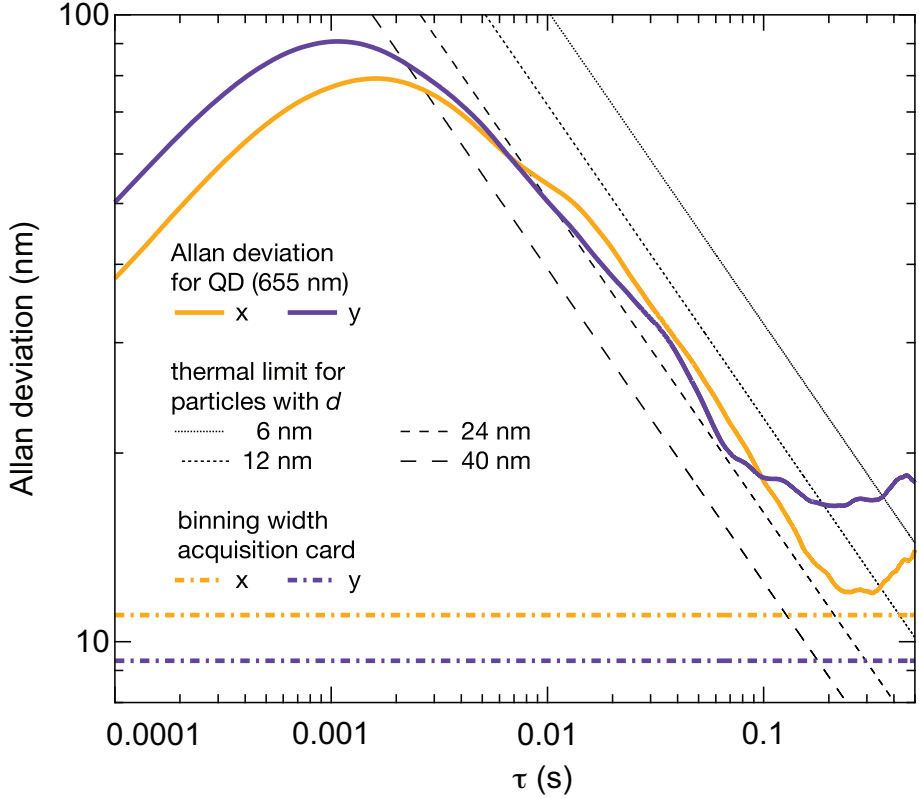} \caption{\label{fig:AllanDevQD}Allan deviation of a 
655~nm QD as a function of measurement time $\tau$ for the two lateral directions
$x$ (orange) and $y$ (purple). The thermal limit is plotted for spherical particles with 
diameters $d$ = 6, 12, 24, and 40 nm, respectively. The optimal measurement time is at
the global minimum for the Allan deviation, $\tau \sim$ 0.3 seconds. 
The noise inherently present in the acquisition card sets a technical limit.}
\end{figure}

The Allan deviation, $\sigma_x\left(\tau\right)$, of an optically trapped 655 nm QD w is plotted 
in Fig. \ref{fig:AllanDevQD}. 
At measurement times below $\tau$ = 0.03 s the positions are oversampled. In the interval
$\tau$ = 0.03 s to $\tau$ = 0.3 s the Allan deviation steadily decreases, thus showing that additional
data acquisition reduces the overall measurement noise, consistent with the positions being
drawn from a Gaussian distribution. Around $\tau$ = 0.3 s the Allan deviation reaches
a minimum, this is the optimal measurement and calibration time for an individual
quantum dot in the optical trap. At larger $\tau$ noise originating from low-frequency drift in the equipment
kicks in and increases the Allan deviation. 

From knowledge of the physical size of a particle the thermal limit for position determination can be 
calculated \cite{fabian2009}. This limit cannot be beaten by any measurement that does not
oversample the signal. The 655 QD depicted in Fig. \ref{fig:AllanDevQD} is elongated ($\sim$ 24 x 40 nm$^2$), and the
straight lines on the figure denote the thermal limits of particles with sizes of 6 nm, 12 nm, 
24 nm, and 40 nm, respectively. The larger the particle, the less its thermal diffusion, the
lower the thermal limit. Also shown in the figure is the limitation originating from the binning of
data on the acquisition card (PCI M-6251, National Instruments). This sets a technical limit to a 
measurable accuracy. The Allan variance of the 655 QD in both lateral directions fall
just above the thermal limit for a 24 nm particle, hence, not much noise is present,
the equipment is performing very well. At present we do not know the orientation of an elongated
QD within the optical trap. It is possible that it orients along the electrical field of the trap 
as is the case for an optically trapped gold nanorod \cite{selhuber-unkel2008}.

The optimal measurement time ($\tau$ = 0.3 seconds) for a trapped colloidal QD
is fairly short in comparison to, e.g., the
optimal measurement time for a micron-sized polystyrene bead ($\tau \sim$ 2-5 seconds) \cite{fabian2009}. 
However, it is significantly longer than the optimal measurement time for a gold nanorod ($\tau \sim$ 0.05 seconds)
\cite{czerwinski2009}, despite the fact that the trap stiffness for a gold nanorod is essentially
identical to the trap stiffness for a colloidal QD \cite{selhuber-unkel2008,jauffred2008}.


\section{Tethered particle motion with quantum dots}
Their small size, low bleaching, and high quantum yield make QDs ideal tracker particles for assays that 
rely on the tethered particle method (TPM). We investigated the 
thermal fluctuation of a QD attached to a 1.36 $\mu$m dsDNA tether. 
The equation of motion of a particle (with low Reynolds number) tethered to a surface
and performing Brownian fluctuations is (in one dimension): 
\begin{equation}
\textbf{F}(t)=\gamma\frac{d\textbf{x}}{dt}+\textbf{F}_{DNA},
\end{equation}
where \textbf{F}$_{DNA}$ is the force exerted by the DNA tether on the particle. The first term on the right hand side depends linearly on the radius, $R$, of the tethered particle. Hence, for small particles the Brownian motion 
is dominated by the macromolecule and not by the tracer particle. The proximity of the tethered particle to 
the anchoring surface affects both the interpretation of the tethered particle's motion and 
the possible conformations of the tethering macromolecule \cite{segall2006}.
The dynamics of the tracer particle does not dominate the dynamics of the tethering
macromolecule when its radius, $R_{QD}$, fulfills
\begin{equation}\label{eq:exc}
R_{QD}< \sqrt{2/3\, N \cdot l_k^2},
\end{equation}
where $l_k$ is the Kuhn length and $N$ the number of segments. The right hand term is derived
as the theoretical root-mean-square displacement (RMS) of the 2D projection \cite{boal2002}. It has a value 294 nm for the 605 nm QD. Hence, with a QD tracer particle we are in the limit where movements of the tracer is dominated by the tethering DNA. 

\subsection{Methods}
The double stranded DNA tether had a length of 1.36 $\mu$m, biotin was specifically attached to one end and
digoxygenin to the other end. Complete information regarding the preparation
of the DNA is found in Reference \cite{thesis}. 
Sample chambers were prepared with an anti-digoxygenin-coating which would bind to the digoxygenin end of the 
DNA tether. After incubation with 0.1 nM DNA in TE  (10 mM Tris-HCl  and 1 mM EDTA) buffer the samples were 
incubated with streptavidin-coated QDs (Invitrogen) in TE buffer that would bind to the biotin-end of the DNA. 
The solution was then interchanged with TE buffer containing 2 mg/ml $\alpha$-casein and the samples were 
sealed with vacuum grease.

The QDs were excited using a Hg lamp.
Images of the samples were collected with a cooled electron-multiplying CCD (Andor Ixon) with a rate of 10/s. 
The image sequences were analyzed with the particle tracker routine Spottracker2D \cite{sage2005}, 
available as plug-in for ImageJ.
The coordinates of the QD were extracted through a tiff-stack. To discriminate multi-tether events only 
symmetric in-plane motion about the anchor point were regarded. 
The second moments of the positions give the co-variance matrix:
\begin{equation}\label{eq:co-variance}
c_{x,y}=N^{-1}\Sigma_{1}^{N}x_{k}y_{k}-\overline{x}\,\overline{y},
\end{equation}
where $x$ and $y$ are the positions in the two directions in the projected plane, $\overline{x}$ and $\overline{y}$ are average displacements around the anchor point, and $N$ is the total number of frames.
If the dynamics of the tracer particle were perfectly symmetric,
the eigenvalues of this matrix $(\lambda_1,\lambda_2)$ would be equal. Our criterion for symmetry was 
$\sqrt{\lambda_{max}/\lambda_{min}}\leq$ 1.1 \cite{han2008}.  

The experimental RMS is a function of the total sampling time or time lag $\tau$:
\begin{equation}\label{eq:RMS}
\unit{RMS}_{\tau}=\sqrt{\langle(x-\overline{x})^2+(y-\overline{y})^2\rangle _{\tau}}.
\end{equation}
For tethered Brownian motion around an anchor point RMS approaches a constant value. 
In addition to the symmetry criterion we had a lower  
limit on the RMS displacement, this is in order to remove those data sets where
multiple (more than two) tethers attach to an individual QD, or the QD is unspecifically
bound to the lower cover slide. In conclusion, we only consider those datasets for which  
RMS $\in\{100\unit{nm}:L\}$, $L$ being the contour length of the tether.

\subsection{Results}
\begin{figure}
\centering
  \includegraphics[width=14cm]{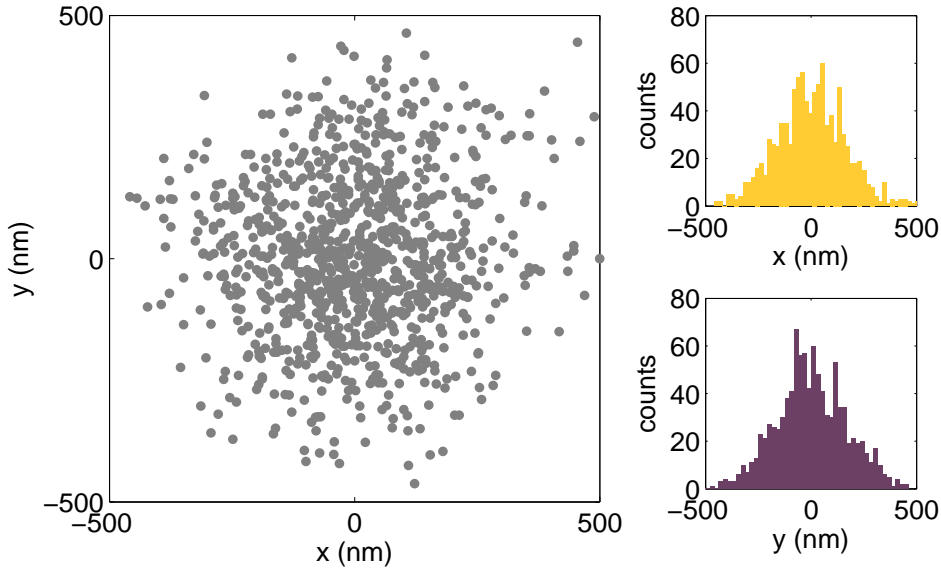}
  \caption{Positions visited by a QD
tethered to a surface with DNA (L $\sim 1.36$ $\mu$m). The 2D projected positions visited 
are plotted to the left. The position distributions of the two lateral directions  are shown to the right.}
\label{fig:qdot}
\end{figure}

One example of a TPM data is shown in Fig. \ref{fig:qdot}. This figure shows a scatter plot of the
positions visited by the tracer QD. The dynamics appears symmetric and indeed fulfills the symmetry
criteria. Also shown are histograms for the positions visited in the two lateral directions.

\begin{figure}
  \begin{center}
  \includegraphics[width=14cm]{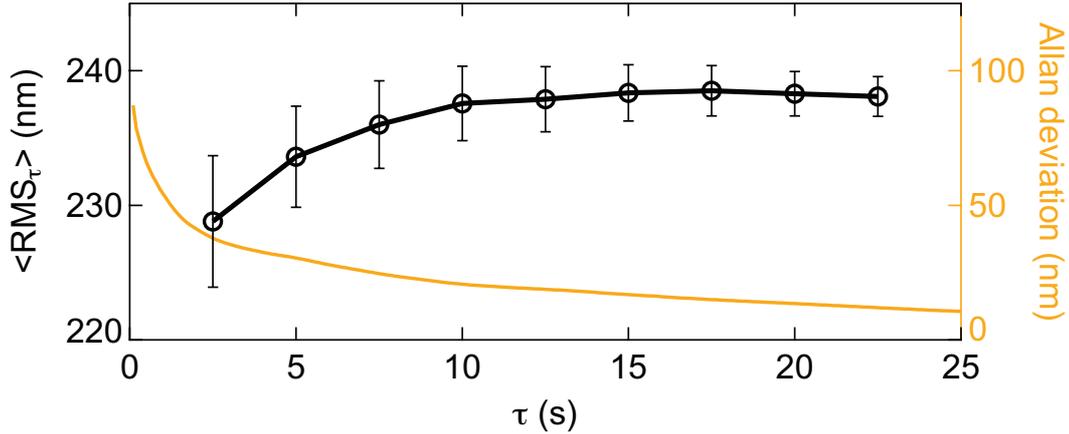}
  \end{center}
  \caption[example] {\label{fig:rms} The black circles (left ordinate axis) show  $\langle$RMS$_{\tau}\rangle$ 
as a function of $\tau$ for TPM of
a QD attached to a DNA tether. Error bars denote one SEM. $\langle$RMS$_{\tau}\rangle$ approaches a constant value
 of 238 nm. The averaged Allan deviation for the same data is plotted on the same graph (yellow curve, right 
ordinate axis). The Allan deviation constantly decreases, no global minimum is obtained within this
measurement. After 10 seconds, the noise is smaller than 10~\% of $\langle$RMS$_{\tau}\rangle$. }
  \end{figure}

Fig.  \ref{fig:rms} shows $\langle$RMS$_{\tau}\rangle$ as a function of time, data originates from
18 individual DNA tethered QDs, all fulfilling the symmetri criteria. $\langle$RMS$_{\tau}\rangle$ reaches 
the expected plateau at 238 nm for $\tau>15$ s. The QDs used in the present study are the smallest
particles every reported as tracer particles for TPM. Their presence has only a small influence
on the dynamics of the DNA tether and hence, they are optimal to use for TPM studies. 
Also, their large photo-stability ensures that they are traceable for a long time. The fluorescence intermittencies of QDs is an issue to consider. However, existing software compensates
for the blinking thus that it becomes unimportant for dynamics on the time-scale of
seconds \cite{dahan2003}.

In principle, Allan variance analysis can be performed on any type of time series. Hence, we also performed
it on the time series of positions visited by the DNA tethered QD. The result is also shown (in yellow)
in Fig. \ref{fig:rms}. Notice that the axes in this figure are linear (axes were logarithmic in
Fig. \ref{fig:AllanDevQD}). The Allan variance decreases with increasing $\tau$ and within our measurement time
it never reaches a global minimum. Hence, with the drift present in this experiment, the system is stable for tens of seconds.

\section{Conclusion}
Quantum dots are optimal handles to use for manipulation and visualization of nano-scale systems. 
With a single infrared tightly focused laser beam one can both optically trap and perform
two-photon excitation of individual colloidal quantum dots. The trapping strength is nearly independent
of quantum dot emission wavelength or its physical size in the range investigated here. Allan variance analysis revealed that the optimal
trapping time of an individual colloidal quantum dot is on the order of 0.3 seconds, which is longer than
the optimal trapping time for a gold nanorod, but shorter than for a $\mu$m sized polystyrene bead trapped
with the same equipment. Individual quantum dots were specifically attached to a DNA tether, which is the
other end was attached to a cover slide, these quantum dots were used to investigate the tethered particle
motion of the combined DNA-quantum dot complex. Due to the small size of the quantum dot, the systems
dynamics was dominated by the DNA tether rather than by the tracer particle. It is, therefore, an ideal system to investigate
dynamics of macromolecules. 

\section{Acknowledgements}
We are grateful to Szabolcs Semsey, ELTE, Budapest, who has prepared the DNA tethers for these experiments.
Also, LBO acknowledges financial support from a University of Copenhagen excellence grant.


\end{document}